\documentclass[journal]{IEEEtran}
\usepackage{amsmath}
\usepackage{amssymb, bm, cite, psfrag, mathtools}

\usepackage{rotating}
\usepackage{dblfloatfix}
\usepackage{algorithm,algpseudocode}
\usepackage{array}
\usepackage{supertabular,booktabs}
\usepackage{ragged2e}
\newcolumntype{P}[1]{>{\centering\hspace{0pt}}p{#1}}
\newcolumntype{M}[1]{>{\centering\hspace{0pt}}m{#1}}
\newcolumntype{L}{>{\centering\arraybackslash}m{3cm}}
\usepackage{changepage}
\usepackage[font=small]{caption}
\usepackage{subcaption}
\usepackage[customcolors,norndcorners]{hf-tikz}
\usepackage{soul}            
\usetikzlibrary{calc}

\usepackage{capt-of}
\usepackage{bbm}
\usepackage{multirow}
\usepackage{xcolor}
\definecolor{Fgreen}{rgb}{0,0.6,0}

\usepackage{footnote}
\makesavenoteenv{tabular}
\makesavenoteenv{table}
\usepackage{etoolbox}
\usepackage{pbox}
\usepackage[hyphens]{url}
\usepackage{cite}
\usepackage{longtable}
\usepackage{tabu}
\usepackage[margin=0.52in,top=0.27in,bottom=0.38in]{geometry}

\usepackage{fixltx2e}
\usepackage{graphicx,etoolbox}
\usepackage{tabu}
\usepackage{booktabs}
\def\PL{\textrm{PL}}
\def\dB{\textrm{dB}}
\def\FSPL{\textrm{FSPL}}
\def\CI{\textrm{CI}}

\def\CIH{\textrm{CIH}}

\def\PLE{\textrm{PLE}}

\def\1m{\textrm{1 m}}
\def\m{\textrm{m}}

\def\LOS{\textrm{LOS}}
\def\NLOS{\textrm{NLOS}}

\newtoggle{conference}
\togglefalse{conference} 
\usepackage{tikz}
\usetikzlibrary{calc}
\begin{document}
\bibliographystyle{IEEEtran}

\title{\LARGE{Study on 3GPP Rural Macrocell Path Loss Models for \\Millimeter Wave Wireless Communications}} 

\author{
George R. MacCartney, Jr.,~\IEEEmembership{Student Member,~IEEE}
and Theodore S. Rappaport,~\IEEEmembership{Fellow,~IEEE}\\
\IEEEauthorblockA{NYU WIRELESS and NYU Tandon School of Engineering, New York University}\vspace{-0.9cm}

\thanks{This material is based upon work published in ``All Things Cellular"~\cite{Mac16c} and is supported by the NYU WIRELESS Industrial Affiliates Program, three National Science Foundation (NSF) Research Grants: 1320472, 1302336, and 1555332, and the GAANN Fellowship Program. Special thanks is given to W. Johnston and B. Ghaffari at the FCC for their assistance in obtaining experimental license number: 1177-EX-ST-2016. The authors also thank S. Sun, Y. Xing, H. Yan, J. Koka, R. Wang, and D. Yu, who helped conduct the propagation measurements. G. R. MacCartney, Jr. (email: gmac@nyu.edu) and T. S. Rappaport (email: tsr@nyu.edu) are with the NYU WIRELESS Research Center, NYU Tandon School of Engineering, New York University, Brooklyn, NY 11201.}
}
\maketitle
\begin{tikzpicture}[remember picture, overlay]
\node[font=\small] at ($(current page.north) + (0,-0.12in)$) {G. R. MacCartney, Jr. and T. S. Rappaport, ``Study on 3GPP Rural Macrocell Path Loss Models for Millimeter Wave Wireless Communications,"};
\node[font=\small] at ($(current page.north) + (0,-0.27in)$)  {in \textit{2017 IEEE International Conference on Communications (ICC)}, Paris, France, May 2017, pp. 1-7.};
\end{tikzpicture}

\begin{abstract}
Little research has been done to reliably model millimeter wave (mmWave) path loss in rural macrocell settings, yet, models have been hastily adopted without substantial empirical evidence. This paper studies past rural macrocell (RMa) path loss models and exposes concerns with the current 3rd Generation Partnership Project (3GPP) TR 38.900 (Release 14) RMa path loss models adopted from the International Telecommunications Union - Radiocommunications (ITU-R) Sector. This paper shows how the 3GPP RMa large-scale path loss models were derived for frequencies below 6 GHz, yet they are being asserted for use up to 30 GHz, even though there has not been sufficient work or published data to support their validity at frequencies above 6 GHz or in the mmWave bands. We present the background of the 3GPP RMa path loss models and their use of odd correction factors not suitable for rural scenarios, and show that the multi-frequency close-in free space reference distance (CI) path loss model is more accurate and reliable than current 3GPP and ITU-R RMa models. Using field data and simulations, we introduce a new close-in free space reference distance with height dependent path loss exponent model (CIH), that predicts rural macrocell path loss using an effective path loss exponent that is a function of base station antenna height. This work shows the CI and CIH models can be used from 500 MHz to 100 GHz for rural mmWave coverage and interference analysis, without any discontinuity at 6 GHz as exists in today's 3GPP and ITU-R RMa models. 
\end{abstract}

\iftoggle{conference}{}{
\begin{IEEEkeywords}
Millimeter wave, mmWave, rural macrocell, RMa, 73 GHz, path loss, channel model, 3GPP, ITU-R, standards.
\end{IEEEkeywords}}

\section{Introduction}\label{sec:intro}
The use of millimeter wave (mmWave) frequencies for fifth-generation (5G) wireless communications offers the promise of multi-gigabit per second data transfers, and vast new consumer and industrial applications~\cite{Rap13a,Pi11a,Boccardi14a,Mac16c}. Thus, many research groups have developed channel models for mmWave frequencies~\cite{METIS15a,Miweba14a,mmMAGIC16a,A5GCM15}. The \textit{3rd Generation Partnership Project (3GPP)}, the global standards body of the wireless industry, released its study on channel models for frequencies above 6 GHz, in 3GPP TR 38.900 V14.2.0 (Release 14)~\cite{3GPP.38.900}. 

The 3GPP channel models in~\cite{3GPP.38.900} were derived from numerous academic and industrial contributions with extensive propagation measurements and ray-tracing simulations~\cite{A5GCM15,Haneda16a,Mac15b,Sun16b}. Scenarios included in the 3GPP TR 38.900 channel model~\cite{3GPP.38.900} are: urban microcell (UMi), urban macrocell (UMa), and indoor hotspot (InH) for office and shopping mall~\cite{Sun16b,Haneda16a}. Rural macrocell (RMa) is a new scenario included in 3GPP TR 38.900~\cite{3GPP.38.900}, that was not included in the 3GPP TR 36.873 (Release 12) LTE channel model study for below 6 GHz~\cite{3GPP.36.873}. The literature shows that the RMa path loss models in~\cite{3GPP.38.900} were adopted from International Telecommunications Union - Radiocommunications (ITU-R) Sector M.2135, but are only specified for frequencies up to 6 GHz in~\cite{ITU-RM.2135}.

This paper describes the development of the existing RMa path loss models in 3GPP and ITU-R~\cite{3GPP.38.900,ITU-RM.2135,Mac16c}, and illuminates model inconsistencies and many questionable empirical correction factors that make no physical sense for a rural setting. A new multi-frequency close-in free space reference distance (CI) path loss model with a path loss exponent (PLE) that is a function of the base station height (CIH) is introduced here, and is analyzed using simulated and measured data. The resulting accurate, physically-based, and simple to use CI and CIH path loss models may be used for coverage and interference analysis for future 5G mmWave networks in macrocellular rural areas with low building density, and may also be adopted by 3GPP and ITU-R.

\section{3GPP RMa Path Loss Models}\label{sec:3GPP_PL}
RMa path loss models are generally used for tall transmitter (TX) heights above 35 meters~\cite{3GPP.38.900,Mac16c}, and are important for predicting the statistical behavior of received signal strength in rural areas. As shown in~\cite{Rap02a,Mac14b,Rap15b,Mac15b}, large-scale path loss is independent of frequency in outdoor macrocell channels, except for the first meter of propagation loss which is a function of the square of the frequency~\cite{Rap02a,Sun16b,Mac14b,Rap15b,Friis46a}. It is noteworthy that path loss models may be developed using either narrowband or wideband signals, since the average received power level at a local area location (in time or space) is independent of bandwidth~\cite{Rap02a}. 

\subsection{3GPP RMa LOS Path Loss Model}\label{sec:3GPP_LOSPL}
The existing 3GPP/ITU-R~\cite{3GPP.38.900,ITU-RM.2135} RMa line-of-sight (LOS) path loss model consists of two sections where the attenuation slope increases beyond a breakpoint distance ($d_{BP}$), as in~\eqref{eq:RMaLOS}~\cite{3GPP.38.900,ITU-RM.2135}:
\begin{align}\label{eq:RMaLOS}
\footnotesize
\begin{split}
PL _1&[\dB] = 20\log_{10}(40\pi \cdot d_{3D} \cdot f_c /3)+\min(0.03h^{1.72},10)\log_{10}(d_{3D})\\
&-\min(0.044h^{1.72},14.77)+0.002\log_{10}(h)d_{3D};\;\;\sigma_{SF}=4\:\dB\\
PL_2&[\dB] = PL_1 (d_{BP})+40\log_{10}(d_{3D}/d_{BP});\;\;\sigma_{SF}=6\:\dB
\end{split}
\end{align}
where $f_c$ is the center frequency in GHz, $d_{3D}$ is the three-dimensional (3D) transmitter-receiver (T-R) separation distance in meters (m), $h$ is the average building height in meters (an odd parameter for RMa LOS), and $d_{BP}$ is the two-dimensional (2D) breakpoint distance along the flat earth in meters. It is worth noting that as the flat earth distance becomes large ($>$ 1 km), the difference between $d_{2D}$ and $d_{3D}$ becomes negligible for typical TX and receiver (RX) heights. $PL_1[\dB]$ is path loss in dB before the breakpoint distance with a shadow fading (SF) standard deviation $\sigma_{SF}=4$ dB and $PL_2[\dB]$ is path loss in dB after the breakpoint distance with a SF standard deviation $\sigma_{SF}=6$ dB. The $PL_2$ equation in~\eqref{eq:RMaLOS} indicates a PLE of 4 after the breakpoint distance, as derived by Bullington for the asymptotic two-ray ground bounce model~\cite{Rap02a,Bullington47a}. The breakpoint distance in~\eqref{eq:RMaLOS} is defined as:
\begin{equation}\label{eq:dbp}
\footnotesize
d_{BP} =  2\pi \cdot h_{BS} \cdot h_{UT} \cdot f_c/c
\end{equation}
where $f_c$ is the center frequency in Hz, $c$ is the speed of light in free space in meters per second, $h_{BS}$ is the base station height in meters, and $h_{UT}$ is the user terminal (UT) height in meters. Table~\ref{tbl:appRange} provides default parameter values and applicability ranges for the 3GPP LOS and non-LOS (NLOS) RMa path loss models~\cite{3GPP.38.900}. 

\begin{table}
	\centering
	\caption{3GPP RMa default path loss model parameters~\cite{3GPP.38.900}.}\label{tbl:appRange}
	\scalebox{0.8}{
		\begin{tabu}{|l|}\hline
			\textbf{RMa LOS Default Values and Applicability Ranges} \\ \specialrule{1.5pt}{0pt}{0pt}
			10 m $\leq d_{2D} \leq d_{BP}$, \\
			$d_{BP} \leq d_{2D} \leq 10\:000$ m,\\
			$h_{BS} = 35$ m, $h_{UT}=1.5$ m, $W=20$ m, $h=5$ m\\
			Applicability ranges: 5 m $\leq h \leq 50$ m; 5 m $\leq W \leq 50$ m; \\
			10 m $ \leq h_{BS} \leq 150$ m; 1 m $\leq h_{UT} \leq 10$ m \\ \hline
			\textbf{RMa NLOS Default Values and Applicability Ranges} \\ \specialrule{1.5pt}{0pt}{0pt}
			10 m $\leq d_{2D} \leq 5\:000$ m, \\
			
			$h_{BS} = 35$ m, $h_{UT}=1.5$ m, $W=20$ m, $h=5$ m\\
			Applicability ranges: 5 m $\leq h \leq 50$ m; 5 m $\leq W\leq 50$ m; \\
			10 m $\leq h_{BS}\leq 150$ m; 1 m $\leq h_{UT}\leq 10$ m \\ \hline
	\end{tabu}}
\end{table}

After a thorough literature review, it was determined that ITU 5D/88-E~\cite{ITU-5D/88-E} is the source of the ITU-R M.2135 RMa LOS path loss model adopted in 3GPP TR 38.900 for frequencies above 6 GHz~\cite{3GPP.38.900}. The LOS model in~\cite{ITU-5D/88-E}, however, was based largely on propagation measurements at 2.6 GHz in 2000 in metropolitan Tokyo (a typical UMi)~\cite{Ichitsubo00a}. The work in~\cite{Ichitsubo00a} developed elaborate correction factors for LOS path loss that used average building height as a physical descriptor to generate models for use in urban cellular prediction at low GHz bands. The RMa LOS path loss model~\eqref{eq:RMaLOS} from ITU-R M.2135~\cite{ITU-RM.2135} and 3GPP~\cite{3GPP.38.900} is similar to the LOS path loss model provided in~\cite{ITU-5D/88-E}. 

The precise RMa LOS path loss model in ITU-R M.2135 and 3GPP 38.900 was not given in~\cite{ITU-5D/88-E,Ichitsubo00a}, or any other published material that we could find, leaving us to conclude that the existing 3GPP RMa LOS path loss model was never confirmed for a rural environment, nor was it confirmed with extensive measurements above 6 GHz or at mmWave bands. At first glance,~\eqref{eq:RMaLOS} is a cumbersome equation that does not have an intuitive physical meaning~\cite{Mac16c}. This is clearly seen by observing the various correction factor ``$\min$" terms which are functions of average building height and are purely curve fitting adjustments. The use of average building height $h$ in~\eqref{eq:RMaLOS} is quite odd, considering that an RMa scenario does not typically have tall buildings. The breakpoint distance used in~\eqref{eq:RMaLOS} and~\eqref{eq:dbp}, however, does have a physical basis (see Bullington~\cite{Bullington47a}), where the asymptotic PLE $n=4$~\cite{Rap02a}. 

\begin{figure}
	\centering
	\includegraphics[width=0.42\textwidth]{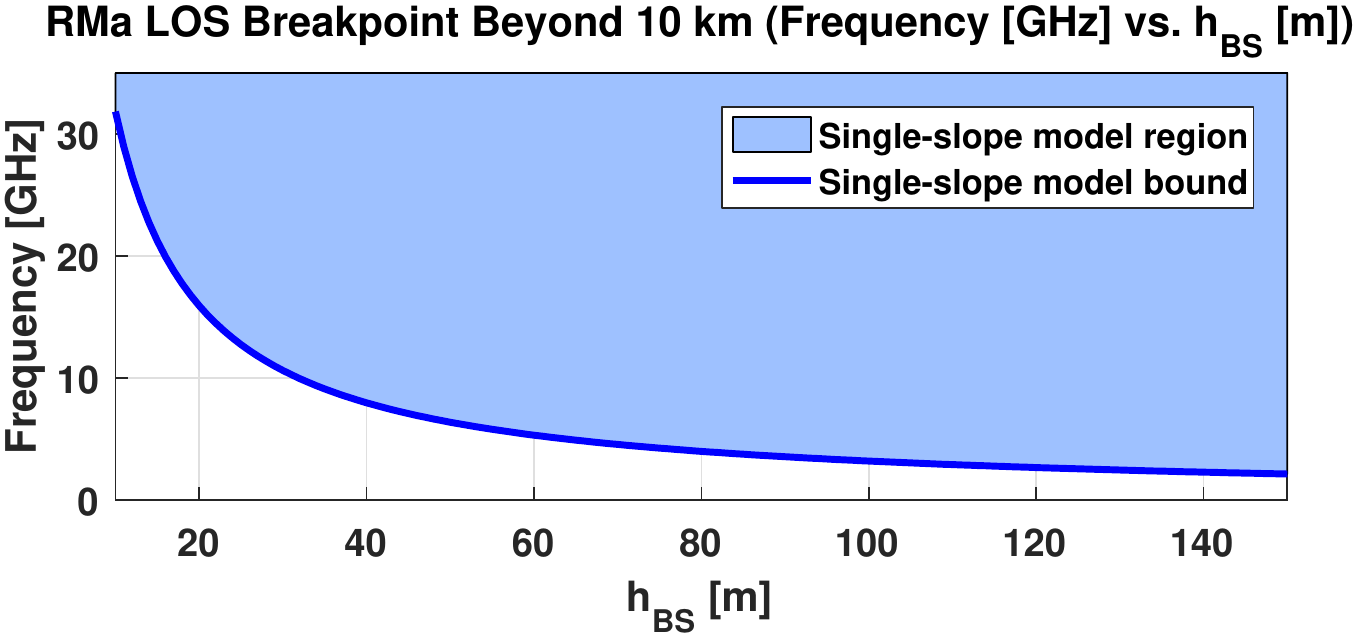}
	\caption{Frequency [GHz] and base station height combinations where the RMa LOS path loss model in~\eqref{eq:RMaLOS} reverts to a single slope model (shaded region) since the breakpoint distance in~\eqref{eq:RMaLOS} and~\eqref{eq:dbp} exceeds 10 km, with a mobile height of 1.5 m.}
	\label{fig:dbp_hbs}
\end{figure}

Although it has a physical basis, the breakpoint distance for RMa LOS path loss in~\eqref{eq:RMaLOS} and~\eqref{eq:dbp} has a surprising frequency limitation. This is easily seen by using the 3GPP default height parameter settings ($h_{BS}=35$ m and $h_{UT}=1.5$ m) given in~\cite{3GPP.38.900} and provided in Table~\ref{tbl:appRange}. With default parameters, the breakpoint distance is greater than the defined maximum distance of the model (10 km) at 9.1 GHz. Fig.~\ref{fig:dbp_hbs} displays the region (shaded) for various base station heights ($h_{BS}$) and frequency combinations where the 3GPP  RMa LOS path loss model breakpoint distance exceeds the maximum 10 km propagation distance model limit. Furthermore, the breakpoint distance is not usable above 32 GHz for \textit{any} $h_{BS}$ value defined in 3GPP  and for a mobile height of 1.5 m. This may be one reason why the RMa LOS path loss model is only applicable up to 30 GHz, according to~\cite{3GPP.38.900}. It should also be noted that the sub-6 GHz WINNER II channel model~\cite{WinnerII} also included a LOS path loss model for RMa, but in a different form than~\eqref{eq:RMaLOS}.

\subsection{3GPP RMa NLOS Path Loss Model}\label{sec:3GPP_NLOSPL}
The RMa NLOS path loss model in 3GPP~\cite{3GPP.38.900} is taken directly from ITU-R~\cite{ITU-RM.2135} and originates from work by Sakagami and Kuboi in~\cite{Sakagami91a}. The empirical model in~\cite{Sakagami91a} was developed from measurements in metropolitan Tokyo in 1991 at 813 MHz and 1433 MHz. Parameters selected for the model include base station antenna height ($h_{b0}$), base station antenna height above the mobile station ($h_{b}$), building height near the base station ($H$), average building height ($<H>$), height of buildings along the street ($h_s$), street width ($W$), and street angle ($\theta$), with all heights and distances in meters and $\theta$ in degrees. A multiple regression analysis was conducted in~\cite{Sakagami91a} to simultaneously solve for nine model coefficients that minimized the variance between the model and data, and resulted in~\cite{Sakagami91a}:
\begin{align}\label{eq:sakagami}
\footnotesize
\begin{split}
PL & _{Sakagami} = 100-7.1\log_{10}(W)+0.023\theta+1.4\log_{10}(h_s)\\
&+6.1\log_{10}(<H>)-(24.37-3.7(H/h_{b0})^2)\log_{10}(h_{b})\\
&+(43.42-3.1\log_{10}(h_{b}))\log_{10}(d)+20.4\log_{10}(f)
\end{split}
\end{align}
where $f$ is frequency in MHz and $d$ is the T-R separation distance in km. The model was extended from 450 MHz to 2200 MHz with an additional frequency extension term not shown here, but which is given in~\cite{Sakagami91a}.

In the literature~\cite{Ohta03a,Kitao08a}, the extended version of the Sakagami model~\eqref{eq:sakagami} replaces all building height terms with the median building height, and substitutes the frequency term with $20\log_{10}(f)$~\cite{Ohta03a,Kitao08a}. An expansion to account for the mobile heights above 1.5 m was also adopted from the Okumura-Hata model~\cite{Hata80a} by: $PL_{Sakagami}-a(h_m)$ where~\cite{Fujii03a,Fujii98a}:
\begin{equation}\label{eq:HataUT}
\footnotesize
a(h_m)=3.2(\log_{10}(11.75h_m))^2-4.97
\end{equation}
and where $h_m$ is the mobile (UT) antenna height in meters. The combination of expansions and extensions of the Sakagami model presented here is what appears in the 3GPP/ITU-R NLOS RMa path loss model~\cite{3GPP.38.900,ITU-RM.2135}:
\begin{align}\label{eq:RMaNLOS}
\footnotesize
\begin{split}
P&L = \max(PL_{RMa-LOS},PL_{RMa-NLOS})\\
P&L _{RMa-NLOS} = 161.04-7.1\log_{10}(W)+7.5\log_{10}(h)\\
&-(24.37-3.7(h/h_{BS})^2)\log_{10}(h_{BS})\\
&+(43.42-3.1\log_{10}(h_{BS}))(\log_{10}(d_{3D})-3)+20\log_{10}(f_c)\\
&-(3.2(\log_{10}(11.75h_{UT}))^2-4.97)\:;\; \sigma_{SF}=8\:\dB
\end{split}
\end{align}
where $W$ is the street width, $h$ is the average building height (combines all building height coefficients~\cite{Kitao08a}), $h_{BS}$ is the base station height, $h_{UT}$ is the mobile (UT) height, and $d_{3D}$ is the 3D T-R separation distance, where all distances and heights are in meters. Additionally, $f_c$ is the center frequency in GHz and the SF standard deviation is set to $\sigma_{SF}=8$ dB. Applicability ranges for the model are provided in Table~\ref{tbl:appRange} as extracted from~\cite{3GPP.38.900}. The ``$\max$" operator in~\eqref{eq:RMaNLOS} acts as a strange mathematical patch and is used to solve a model artifact where predicted NLOS signal strength is much stronger close-in (say within a few hundred meters) than what theoretical free space path loss (FSPL) predicts, something that is nonsensical and defies physics~\cite{Sun16b}. The patch ensures that the estimated NLOS path loss is always greater than or equal to the equivalent LOS path loss for the same T-R distance. This problem was shown to exist in other path loss models, leading to the optional CI-based path loss models in 3GPP~\cite{3GPP.38.900,Sun16b,A5GCM15}. A footnote for~\eqref{eq:RMaNLOS} in~\cite{3GPP.38.900} specifies the applicable frequency range as 0.8 GHz $< f_c <$ 30 GHz for RMa, although evidence presented herein suggests there is little work to justify this model at frequencies above 6 GHz. 

There are three differences between the extended Sakagami~\eqref{eq:sakagami} and the 3GPP/ITU-R~\eqref{eq:RMaNLOS} NLOS path loss models. The first difference is the removal of the street angle term: $0.023\theta$. The second change is the modification of the first term in~\eqref{eq:RMaNLOS} from 100 to 161.04, since the 3GPP model is in units of GHz rather than MHz (FSPL difference at 1 m between 1 MHz and 1 GHz is 60 dB). The additional 1.04 dB difference was not explained in the standards or literature. The third difference is the addition of ``-3" in $(\log_{10}(d_{3D})-3)$, to account for the fact that $d_{3D}$ is in meters instead of km, as it is in~\cite{Sakagami91a} ($\log_{10}(1000) = 3$). 

The use of~\eqref{eq:RMaNLOS} as a rural scenario path loss formula is questionable, based on the fact that the measurements were made at 813 MHz and 1433 MHz~\cite{Sakagami91a}, much lower than the 30 GHz upper bound specified in 3GPP~\cite{3GPP.38.900}. An extension for use up to 2200 MHz was provided in~\cite{Sakagami91a}, but was not included in 3GPP or ITU-R~\cite{3GPP.38.900,ITU-RM.2135}. It is worth noting that others in the literature have attempted to extend the Sakagami model based on measurements up to 8.45 GHz in urban environments, yet this introduced more correction factors and evolutions of the Sakagami model, that can only be applied in urban environments~\cite{Kitao08a}. While physical parameters such as average building height and street width make sense for modeling urban scenarios, they are not used in any of the other 3GPP (e.g. UMi or UMa) path loss models~\cite{3GPP.38.900}. It is concerning that the NLOS RMa path loss model in 3GPP is strictly based on urban measurements. The mishandling of this urban model is presented in~\cite{Omote16a,Mac16c} which show that the extended Sakagami path loss prediction formula is unsuitable for areas with extremely low average building height. Furthermore, the authors in~\cite{Omote16a} conclude that~\eqref{eq:RMaNLOS} is only applicable for areas with average building heights greater than 5 m. 

The only effort we could find to validate~\eqref{eq:RMaNLOS} in~\cite{3GPP.38.900} was from a small measurement campaign at 24 GHz conducted over limited 2D T-R separation distances between 200 to 500 m~\cite{3GPPTDOC164975}, even though~\eqref{eq:RMaNLOS} is specified for 2D distances up to 5 km for NLOS (and the original Sakagami model in~\cite{Sakagami91a} was valid up to 10 km). The work in~\cite{3GPPTDOC164975} indicates a reasonable match between the measurements and model from 200 to 500 m, but LOS and NLOS path loss data were combined together and best-fit indicators (e.g. RMSE) were not provided, causing one to question the validation. The WINNER II channel model also included a NLOS RMa path loss model~\cite{WinnerII}, but takes on a different form than that in~\eqref{eq:RMaNLOS}. Due to the limited empirical validation of RMa path loss above 6 GHz, we conducted a measurement campaign at 73 GHz in~\cite{Mac16c}.

\section{73 GHz RMa Measurements}\label{sec:meas}
Path loss measurements were conducted in Riner and Christiansburg, Virginia, rural towns in southwest Virginia, USA, at the 73 GHz mmWave band with a narrowband (CW) signal, as described in~\cite{Mac16c}. The measurements were performed with high-gain narrowbeam horn antennas, with a maximum measurable path loss of 190 dB and with 11.7 dBW of effective isotropic radiated power (EIRP). The RMa measurements were made over a two day period in clear weather using a receiver measurement van, with the RX antenna fixed on a tripod outside of the van at an average height between 1.6 m and 2 m above the ground, along country roads and streets near rural homes and businesses. Measurable signal was detected at 14 LOS and 17 NLOS locations. The 2D T-R separation distance for LOS locations ranged from 33 m to 10.8 km, and from 3.4 km to 10.6 km for NLOS locations. The average TX height for all measurements was approximately 110 meters, with additional details provided in~\cite{Mac16c}.

\section{Novel RMa Path Loss Models and Simulations}\label{sec:NewRMa}
Two alternatives to the existing 3GPP RMa path loss models in~\eqref{eq:RMaLOS} and~\eqref{eq:RMaNLOS} are now proposed, based on the optional path loss models in 3GPP~\cite{3GPP.38.900,Haneda16a} and as found in~\cite{Sun16b,Mac15b,Rap15b}. A CI RMa path loss model is proposed, using a 1 m reference distance. A new model with a TX height dependent PLE and a 1 m reference distance (CIH model) is also presented. Both the CI and CIH models are shown to have a solid physical basis, are proven to be accurate and reliable, and are easy to understand and apply, especially in the mmWave bands~\cite{Sun16b,Rap15b,Mac15b,Haneda16a,A5GCM15}. CI models in earlier work have also proven to be stable and accurate when predicting path loss for scenarios and distances outside the scope of the original measurements~\cite{Sun16b}. Such models allow for a single equation and few parameters when predicting path loss over a broad range of frequencies, from microwave to mmWave.

\subsection{CI Path Loss Model}
The simplest form of the CI model, with a 1 m free space reference distance ($d_0$)~\cite{Andersen17a,Andersen95a}, was adopted as an optional model for UMa, UMi, and InH scenarios in 3GPP~\cite{3GPP.38.900}, based on numerous experiments at mmWaves~\cite{Rap15b,Haneda16a,A5GCM15,Mac15b,Sun16b}. Thus, it would also seem reasonable to consider a CI option for the RMa scenario in 3GPP and ITU-R. We show subsequently from the measured data that indeed the CI model is a good fit for predicting RMa path loss.

The general expression for the CI path loss model is:
\begin{equation}\label{eq:CIgen}
\footnotesize
	\begin{split}
		\PL^{\CI}(f_c,d)[\dB]&=\FSPL(f_c, d_0)[\dB]+10n\log_{10}\left(\frac{d}{d_0}\right)+\chi_{\sigma};\\
		&~\text{where}~d\geq d_0\text{ and } d_0 = 1\;\m
	\end{split}
\end{equation}
where $d$ (usually 3D distance) is the T-R separation in meters between the TX and RX, $d_0$ is the close-in free space reference distance set to 1 m, $n$ represents the PLE~\cite{Rap15b,Rap02a}, and $f_c$ is frequency in GHz. The SF is represented by a zero-mean Gaussian random variable $\chi_\sigma$ with standard deviation $\sigma$ in dB~\cite{Rap15b}. For large T-R separation distances (several km) such as in the RMa scenario, the distance $d$ may be represented by the 2D or 3D distance, as the difference is minuscule. 

The first term after the equality sign in~\eqref{eq:CIgen} models frequency-dependent path loss up to the close-in reference distance $d_0=1$ m~\cite{Rap02a}, and is equivalent to Friis' FSPL~\cite{Friis46a,Rap02a}: $\FSPL(f_c,1\text{ m})[\dB]=32.4+20\log_{10}(f_c)$ where $f_c$ is the center frequency in GHz, $c$ is the speed of light in free space or air ($3\times 10^8$ m/s), and 32.4 dB is the FSPL at 1 m at 1 GHz, which yields:
\begin{equation}\label{eq:CI2}
\footnotesize
\begin{split}
\PL^{\CI}(f_c,d)[\dB]=\:&32.4+10n\log_{10}(d)+20\log_{10}(f_c)+\chi_{\sigma};~\text{where}~d\geq \text{1 m}
\end{split}
\end{equation}
Setting $d_0=1$ m provides a standardized and universal modeling approach for path loss comparison using a single parameter, the PLE~\cite{A5GCM15,Rap15b,Rap02a,Haneda16a,Sun16b}. The CI model for RMa~\eqref{eq:CI2} requires only a single parameter $n$ -- the PLE -- to describe the mean path loss over distance for a wide range of mmWave bands, as shown in 3GPP (optional model~\cite{3GPP.38.900})~\cite{Rap15b,Sun16b}. The use of $d_0=1$ m in~\eqref{eq:CI2} makes physical sense because there are clearly no obstructions in the first meter of propagation from a base station antenna, and it accurately models the frequency dependency of propagation in outdoor channels over a vast span of frequencies~\cite{Sun16b,Rap15b}. 

\subsection{CIH Path Loss Model}
When considering the existing RMa path loss models in 3GPP and ITU-R, there are model parameters such as building height and street width that do not make physical sense, yet others, such as TX and RX height above ground, are expected to impact path loss in rural scenarios. Since the current 3GPP RMa path loss models consider TX heights as low as 10 m and as tall as 150 m, this parameter clearly has much greater range and physical significance than other model parameters (a simple simulation below confirms this). The RX height in the rural scenario, as specified in 3GPP, ranges only 1.5 m to 10 m, which seems negligible when considering T-R separation distances of many kilometers. Therefore, we chose TX height as the lone significant, environment parameter from~\eqref{eq:RMaLOS} and~\eqref{eq:RMaNLOS} to include for RMa path loss modeling.

\begin{figure}
	\centering
	\includegraphics[width=0.46\textwidth]{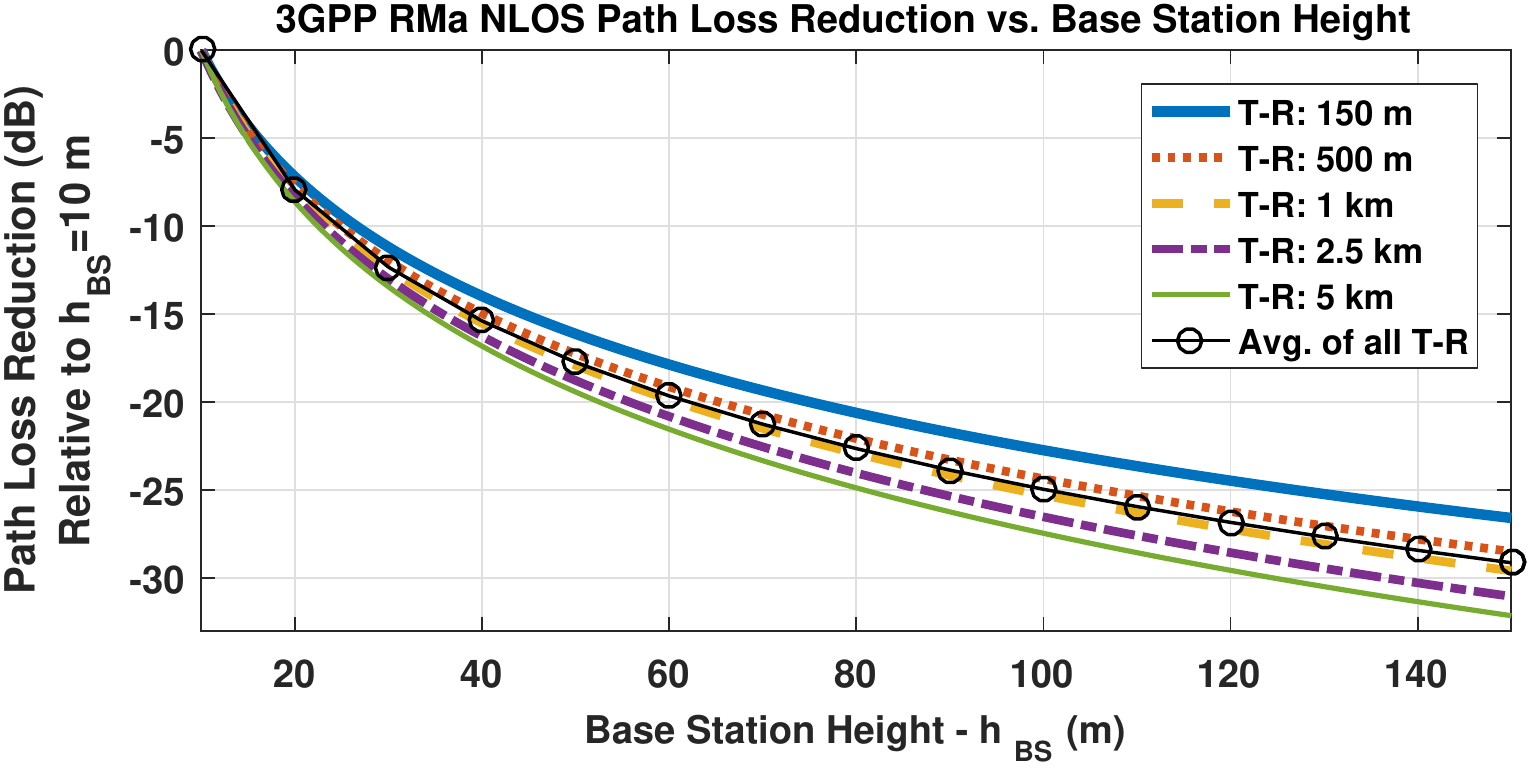}
	\caption{Relationship between TX base station height and decrease in NLOS path loss for T-R separation distances from 150 m to 5 km and the average over all T-R distances, for the 3GPP RMa NLOS path loss model~\eqref{eq:RMaNLOS}. The mobile height ($h_{UT}$) is 1.5 m.}\label{fig:3GPPTXheight}	
\end{figure}

Fig.~\ref{fig:3GPPTXheight} shows the effect of base station height ($h_{BS}$) on NLOS path loss from~\eqref{eq:RMaNLOS} for a wide range of 3D T-R separation distances (150 m, 500 m, 1 km, 2.5 km, and 5 km), and the average effect over all distances.   
The reductions in path loss shown in Fig.~\ref{fig:3GPPTXheight} for the 3GPP NLOS RMa model are independent of frequency (base station height corrections factors in~\eqref{eq:RMaNLOS} do not include frequency), and show that by increasing the TX height from 10 m to 150 m in~\eqref{eq:RMaNLOS}, the path loss is effectively reduced by approximately 26 dB and 32 dB for T-R separation distances of 150 m and 5 km, respectively. This difference in path loss shows a potential one thousand times improvement in received power at any frequency at the RX when increasing the TX antenna height from 10 m to 150 m. Therefore, TX height was deemed an important modeling parameter for RMa path loss estimation.

Here we extend the CI model to include various base station heights (CIH model), such that the model remains physically grounded to FSPL at a close-in reference distance, but also reliably models the PLE dependence on base station height. The CIH model was inspired by the CIF model introduced in~\cite{Mac15b,Sun16b}, which was shown to model the monotonic frequency dependence of path loss for various indoor channels that repeatedly exhibited increased path loss as frequency increased. The CIF model incorporated the PLE to be a function of frequency to account for frequency dependence empirically observed in the environment, while retaining a fundamental physical basis of frequency dependence due to Friis' equation at close-in distances~\cite{Sun16b,Rap15b,Mac15b}.

Until this work, the literature has dealt with many ad hoc and complicated, non-physical correction factors to deal with the impact of TX antenna height~\cite{Kitao08a,Sakagami91a,3GPP.38.900,Ohta03a,Fujii98a,Ichitsubo00a}. By incorporating the TX height as an adjustment to the PLE, we postulated that it could be possible to model secondary path loss effects due to antenna height, just as the CIF model captured secondary frequency-dependent effects, while retaining the physics of the primary frequency dependence of FSPL at close-in distances.

The CIH model takes on a similar form as the CIF model~\cite{Mac15b} and is given for $d_0=1$ m:
\begin{equation}\label{eq:CIH3GPP}
\footnotesize
\begin{split}
\PL^{\CIH}&(f_c,d,h_{BS})[\dB]=32.4+20\log_{10}(f_c)\\
&+10n\left(1+b_{tx}\left(\frac{h_{BS}-h_{B0}}{h_{B0}}\right)\right)\log_{10}(d)+\chi_{\sigma};\\
&\text{where}~d\geq \text{1 m, and }h_{B0}= \text{avg. BS height}\\
\end{split}
\end{equation}
where $h_{BS}$ is the RMa base station height in meters, and $h_{B0}$ is the default base station height or the average of transmitter heights from a measurement set. The distance dependence of path loss is denoted by $n$ (similar to the PLE in the CI model), and $b_{tx}$ is a model parameter that is an optimized weighting factor that scales the parameter $n$ as a function of the base station height relative to the average base station height $h_{B0}$. Similar to the CIF model, the CIH model reduces to the CI model when $b_{tx} = 0$ (no dependence on base station height beyond the first meter of free space propagation), or when $h_{BS}$ = $h_{B0}$. The effective PLE ($\textrm{PLE}_{eff}$) is the PLE that results when $n$ is scaled by $b_{tx}$ and the base station heights in~\eqref{eq:CIH3GPP}, such that: $\textrm{PLE}_{eff}=n\cdot\left(1+b_{tx}\left(\frac{h_{BS}-h_{B0}}{h_{B0}}\right)\right)$. The effective PLE is similar to the \textit{equivalent PLE} defined in~\cite{Azevedo15a}.

\subsection{3GPP RMa Monte Carlo Simulations}
To compare the modeling accuracy of the CI and CIH RMa path loss models~\eqref{eq:CI2},~\eqref{eq:CIH3GPP} with the current 3GPP LOS~\eqref{eq:RMaLOS} and NLOS~\eqref{eq:RMaNLOS} RMa path loss models in~\cite{3GPP.38.900}, we used the default parameters in Table~\ref{tbl:appRange} and performed Monte Carlo simulations for two cases. 
\subsubsection{Case One -- Simulation For 3GPP Default Parameters}
50,000 random path loss samples were generated from~\eqref{eq:RMaLOS} and~\eqref{eq:RMaNLOS}, for the following frequencies: 1, 2, 6, 15, 28, 38, 60, 73, and 100 GHz, resulting in 450,000 samples each (50,000 samples $\times$ 9 frequencies) for LOS and NLOS. Frequencies below and above 6 GHz are used for simulations since the applicable path loss model frequency range for a majority of scenarios in~\cite{3GPP.38.900} is from 0.8 GHz to 100 GHz. Each path loss sample was randomly generated for a 2D T-R separation -- from which 3D distances were calculated with trigonometry -- distance ranging between 10 m and 10 km for LOS and between 10 m and 5 km in NLOS, and with corresponding SF values (in dB) from~\eqref{eq:RMaLOS} and~\eqref{eq:RMaNLOS}. From the simulated 3GPP path loss samples for each environment, CI model parameters with the best fit to the data were derived that resulted in the minimum root mean squared error (RMSE) between the model and the data. The CI models derived from simulated LOS ($\PL^{\CI\text{-3GPP}}_{\LOS}$) and NLOS ($\PL^{\CI\text{-3GPP}}_{\NLOS}$) path loss samples are as follows: 
\begin{equation}\label{eq:CILOS3GPP}
\footnotesize
\begin{split}
\PL^{\CI\text{-3GPP}}_{\LOS}&(f_c,d)[\dB]=32.4+23.1\log_{10}(d)+20\log_{10}(f_c)+\chi_{\sigma_{\LOS}};\\
&\text{where}~d\geq 1~\m,\text{ and }\sigma_{\LOS}=5.9~\dB\\
\end{split}
\end{equation}

\begin{equation}\label{eq:CINLOS3GPP}
\footnotesize
\begin{split}
\PL^{\CI\text{-3GPP}}_{\NLOS}&(f_c,d)[\dB]=32.4+30.4\log_{10}(d)+20\log_{10}(f_c)+\chi_{\sigma_{\NLOS}};\\
&\text{where}~d\geq 1~\m,\text{ and }\sigma_{\NLOS}=8.2~\dB\\
\end{split}
\end{equation}
Both the LOS and NLOS CI models in~\eqref{eq:CILOS3GPP} and~\eqref{eq:CINLOS3GPP} emphatically show that the complicated and questionable 3GPP/ITU-R RMa path loss models in~\eqref{eq:RMaLOS} and~\eqref{eq:RMaNLOS} can be reformulated into succinct and easy to understand equations with nearly identical performance in RMSE. A SF standard deviation of 5.9 dB was determined when using a single PLE parameter for the CI LOS model as compared to 4 to 6 dB in the existing 3GPP LOS model in~\eqref{eq:RMaLOS}. Similarly, a SF standard deviation of 8.2 dB was determined when using the simple single-parameter CI model for NLOS RMa as compared to the NLOS 3GPP RMa model SF standard deviation of 8.0 dB in~\eqref{eq:RMaNLOS}. The CI models in~\eqref{eq:CILOS3GPP} and~\eqref{eq:CINLOS3GPP} exhibit the physics of free space transmission in the first meter of propagation, and show that RMa path loss can be modeled by a simple parameter (PLE), which is independent of frequency for all distances beyond one meter.

\subsubsection{Case Two -- Simulation For 3GPP Default Parameters with Varying Base Station Heights}
LOS and NLOS 3GPP models in~\eqref{eq:RMaLOS} and~\eqref{eq:RMaNLOS} were simulated again, but for base station height ($h_{BS}$) variations from 10 m to 150 m in 5 m increments, and across the same frequencies as in \textit{Case One}, resulting in 13,050,000 samples each (50,000 samples $\times$ 9 frequencies $\times$ 29 base station heights), for LOS and NLOS. From these samples, the best-fit CIH path loss model parameters~\eqref{eq:CIH3GPP} were derived that minimized the RMSE between the model and simulated data. In order to match the 3GPP model, $h_{B0}$ was set to 35 m (default TX height from Table~\ref{tbl:appRange}). The best fit CIH LOS ($\PL^{\CIH\text{-3GPP}}_{\LOS}$) and NLOS ($\PL^{\CIH\text{-3GPP}}_{\NLOS}$) path loss models derived are:
\begin{equation}\label{eq:CIH3GPP_LOSsim}
\footnotesize
\begin{split}
\PL^{\CIH\text{-3GPP}}_{\LOS}&(f_c,d,h_{BS})[\dB]=32.4+20\log_{10}(f_c)\\
&+23.1\left(1-0.006\left(\frac{h_{BS}-35}{35}\right)\right)\log_{10}(d)+\chi_{\sigma_{\LOS}};\\
&\text{where}~d\geq 1~\m,\text{ and }\sigma_{\LOS}=5.6~\dB
\end{split}
\end{equation}
\begin{equation}\label{eq:CIH3GPP_NLOSsim}
\footnotesize
\begin{split}
\PL^{\CIH\text{-3GPP}}_{\NLOS}&(f_c,d,h_{BS})[\dB]=32.4+20\log_{10}(f_c)\\
&+30.7\left(1-0.06\left(\frac{h_{BS}-35}{35}\right)\right)\log_{10}(d)+\chi_{\sigma_{\NLOS}};\\
&\text{where}~d\geq 1~\m,\text{ and }\sigma_{\NLOS}=8.7~\dB
\end{split}
\end{equation} 

The best-fit LOS CIH path loss model in~\eqref{eq:CIH3GPP_LOSsim}, shows that path loss in LOS is slightly dependent on the base station height, as the value $b_{tx}=-0.006$ shows a minuscule decrease in the effective PLE (the coefficient before the $\log_{10}(d)$ term in~\eqref{eq:CIH3GPP_LOSsim}) as the base station height is increased. This observation is likely due to~\eqref{eq:RMaLOS} not explicitly having a correction factor for the TX height, although the breakpoint distance~\eqref{eq:dbp} is a function of the TX height ($h_{BS}$). The SF of 5.6 dB compared to 4 to 6 dB from the 3GPP model in~\eqref{eq:RMaLOS}, indicates that the CIH model fits the simulated data well. 

The dependence of base station height is more noticeable for the NLOS CIH path loss model in~\eqref{eq:CIH3GPP_NLOSsim} with $b_{tx}=-0.06$, compared to the LOS CIH path loss model with $b_{tx}=-0.006$ in~\eqref{eq:CIH3GPP_LOSsim}. This demonstrates that as the base station height increases from 10 m to 150 m, the effective PLE reduces from 3.2 to 2.5, which is 7 dB per decade of distance, resulting in significantly different path losses at long-range distances. The RMSE for the CIH model from simulated data is 8.7 dB, similar to the 8.0 dB SF standard deviation from 3GPP~\cite{3GPP.38.900}. The similar RMSE values indicate that it is reasonable to use the simple CIH model for estimating NLOS RMa path loss as a function of base station height. The comparable RMSE performances between the CI and CIH models \eqref{eq:CILOS3GPP}-\eqref{eq:CIH3GPP_NLOSsim} fit to simulated data versus the 3GPP models in~\eqref{eq:RMaLOS} and~\eqref{eq:RMaNLOS}, show that the CI and CIH models are reliable in predicting RMa path loss, and have much simpler forms and fewer parameters than 3GPP/ITU-R~\cite{3GPP.38.900,ITU-RM.2135}. The best-fit CI and CIH model parameters and corresponding equation numbers from the simulated data are provided in Table~\ref{tbl:params}.

\section{Empirically-based RMa Path Loss models}\label{sec:EmpModel}
Path loss values at 73 GHz for LOS and NLOS were calculated from measurements~\cite{Mac16c} described in Section~\ref{sec:meas} and were used as sample data to derive the best-fit model parameters for the CI and CIH path loss models (See~\cite{Mac15b,Sun16b} for closed-form best-fit optimization approach). 

\subsection{Empirical CI Model Results}
\begin{figure}
	\centering
	\includegraphics[width=0.44\textwidth]{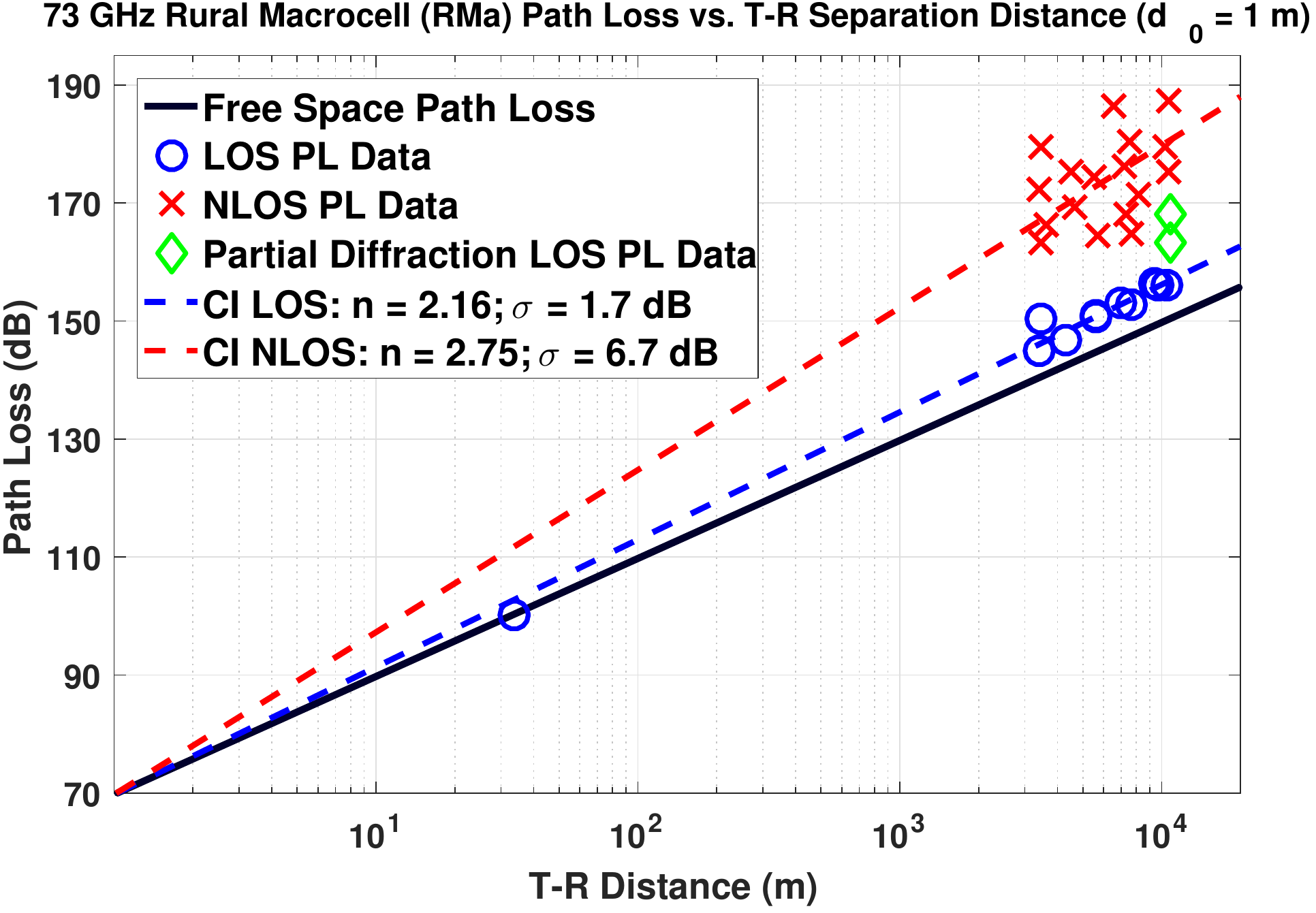}
	\caption{Measured 73 GHz RMa path loss vs. T-R separation distance with LOS and NLOS CI path loss models with a 1 m reference distance.}
	\label{fig:RMaPL1m}
\end{figure}

Measured path loss data and the corresponding CI models are compared with free space in Fig.~\ref{fig:RMaPL1m}. Local time and small-scale spatial averaging were performed over a few seconds to record the received power levels, with small fluctuations in received power ranging from fractions of a dB about the mean power in LOS locations and approximately 3-5 dB in NLOS locations (due to small-scale variations and foliage movement caused by wind). In Fig.~\ref{fig:RMaPL1m}, blue circles represent the measured LOS path loss values, red crosses represent measured NLOS path loss data, and green diamonds denote measured LOS data with partial diffraction (see~\cite{Mac16c}). 

A LOS PLE of 2.16 was calculated from the 73 GHz measurement data with a 110 m TX height, and is very close to FSPL (PLE of 2~\cite{Friis46a,Rap15b,Rap02a}). The distances up to and beyond 10 km with which the LOS links were made is quite remarkable, especially given the close agreement with FSPL and the very small transmit EIRP of 41.7 dBm. The green diamond LOS path loss values include diffraction loss and were not used for the CI model derivation, but are shown to indicate the impact of diffraction edges close to the transmitter (15-20 dB additional loss over long range distances, see~\cite{Mac16c}).

Measured data in clear weather RMa NLOS provided a PLE of 2.75, which is lower than NLOS UMi and UMa mmWave PLEs reported in the literature (between 2.9 and 3.2~\cite{Sun16b,Rap15b}), indicating a slight improvement in received signal level over distance, due to taller base station heights and the lack of tall building obstructions in RMa scenarios. Tall rural cell sites are called ``boomer cells", as they can increase the coverage in a rural area beyond a typical 2-3 mile radius, by using very tall base stations. 

The CI path loss models are provided here for the RMa LOS and NLOS scenarios derived from the measured data at 73 GHz with a 110 m TX height, and are much simpler with comparable accuracy to the existing 3GPP/ITU-R RMa path loss models given in~\eqref{eq:RMaLOS} and~\eqref{eq:RMaNLOS}:
\begin{equation}\label{eq:CILOSmeas}
\footnotesize
\begin{split}
\PL^{\CI\text{-RMa}}_{\LOS}(f_c,d)[\dB]&=32.4+21.6\log_{10}(d)+20\log_{10}(f_c)+\chi_{\sigma_{\LOS}};\\
&\text{where}~d\geq 1~\m\text{, and }\sigma_{\LOS}=1.7\; \dB
\end{split}
\end{equation}
\begin{equation}\label{eq:CINLOSmeas}
\footnotesize
\begin{split}
\PL^{\CI\text{-RMa}}_{\NLOS}(f_c,d)[\dB]&=32.4+27.5\log_{10}(d)+20\log_{10}(f_c)+\chi_{\sigma_{\NLOS}};\\
&\text{where}~d\geq 1~\m\text{, and }\sigma_{\NLOS}=6.7\; \dB
\end{split}
\end{equation}
Similar studies in UMa showed that the PLE is not a function of frequency in the CI model beyond the first meter of propagation~\cite{Sun16b,Haneda16a}. The same CI-based equations are used for the optional 3GPP path loss models for UMi, UMa, and InH~\cite{3GPP.38.900}. The LOS and NLOS RMa CI path loss models given here are also valid up to and beyond 10 km based on the measurement range, well in agreement with the existing 10 km and 5 km distance limits in 3GPP~\cite{3GPP.38.900} for LOS and NLOS environments, respectively. 

\subsection{Empirical CIH Model Results}
Path loss data from multiple TX heights can be used to derive optimal CIH path loss model parameters. Since path loss data for measurements described in Section~\ref{sec:meas} were obtained only for a single TX height (110 m), we would expect that the CIH model~\eqref{eq:CIH3GPP} for a fixed TX height would revert to the CI model~\eqref{eq:CI2}. Therefore, we used the LOS and NLOS CI models~\eqref{eq:CILOSmeas}-\eqref{eq:CINLOSmeas} derived from the 73 GHz measurements and set them equal to the CIH models~\eqref{eq:CIH3GPP_LOSsim}-\eqref{eq:CIH3GPP_NLOSsim} that were derived from 3GPP simulated data, keeping $h_{B0}=35$ m, and setting $h_{BS}=110$ m (the base station height of the 73 GHz measurements). From this equality, it is possible to solve for $b_{tx}$ in order to determine the base station height dependence on path loss. This resulted in $b_{tx}$ values of -0.03 and -0.049 in LOS and NLOS, respectively, and the following LOS ($\PL^{\CIH\text{-RMa}}_{\LOS}$) and NLOS ($\PL^{\CIH\text{-RMa}}_{\NLOS}$) empirically-based CIH RMa path loss models:
\begin{equation}\label{eq:CIHLOSmeas}
\footnotesize
\begin{split}
\PL^{\CIH\text{-RMa}}_{\LOS}&(f_c,d,h_{BS})[\dB]=32.4+20\log_{10}(f_c)\\
&+23.1\left(1-0.03\left(\frac{h_{BS}-35}{35}\right)\right)\log_{10}(d)+\chi_{\sigma_{\LOS}};\\
&\text{where}~d\geq 1~\m,\text{ and }\sigma_{\LOS}=1.7~\dB\\
\end{split}
\end{equation}
\begin{equation}\label{eq:CIHNLOSmeas}
\footnotesize
\begin{split}
\PL^{\CIH\text{-RMa}}_{\NLOS}&(f_c,d,h_{BS})[\dB]=32.4+20\log_{10}(f_c)\\
&+30.7\left(1-0.049\left(\frac{h_{BS}-35}{35}\right)\right)\log_{10}(d)+\chi_{\sigma_{\NLOS}};\\
&\text{where}~d\geq 1~\m,\text{ and }\sigma_{\NLOS}=6.7~\dB\\
\end{split}
\end{equation}
By setting $h_{BS}=110$ m, the RMa models in~\eqref{eq:CIHLOSmeas} and~\eqref{eq:CIHNLOSmeas} revert to the RMa LOS and NLOS CI path loss models in~\eqref{eq:CILOSmeas} and~\eqref{eq:CINLOSmeas} with effective PLEs of 2.16 and 2.75, respectively. The $b_{tx}$ values of -0.03 and -0.049 in LOS and NLOS, respectively, demonstrate the same trend as the CIH models from simulated data in~\eqref{eq:CIH3GPP_LOSsim} and~\eqref{eq:CIH3GPP_NLOSsim} such that the base station height has a considerable influence on the effective PLE. Negative $b_{tx}$ values reveal that the effective PLE decreases as the TX height increases, and this intuitively makes sense since higher base stations would result in fewer building and terrain obstructions, compared to transmitters closer to the ground. Table~\ref{tbl:params} provides the empirical CI and CIH RMa path loss model parameters in LOS and NLOS for comparison with the path loss model parameters derived from simulated data.

\begin{figure}
	\centering
	\includegraphics[width=0.46\textwidth]{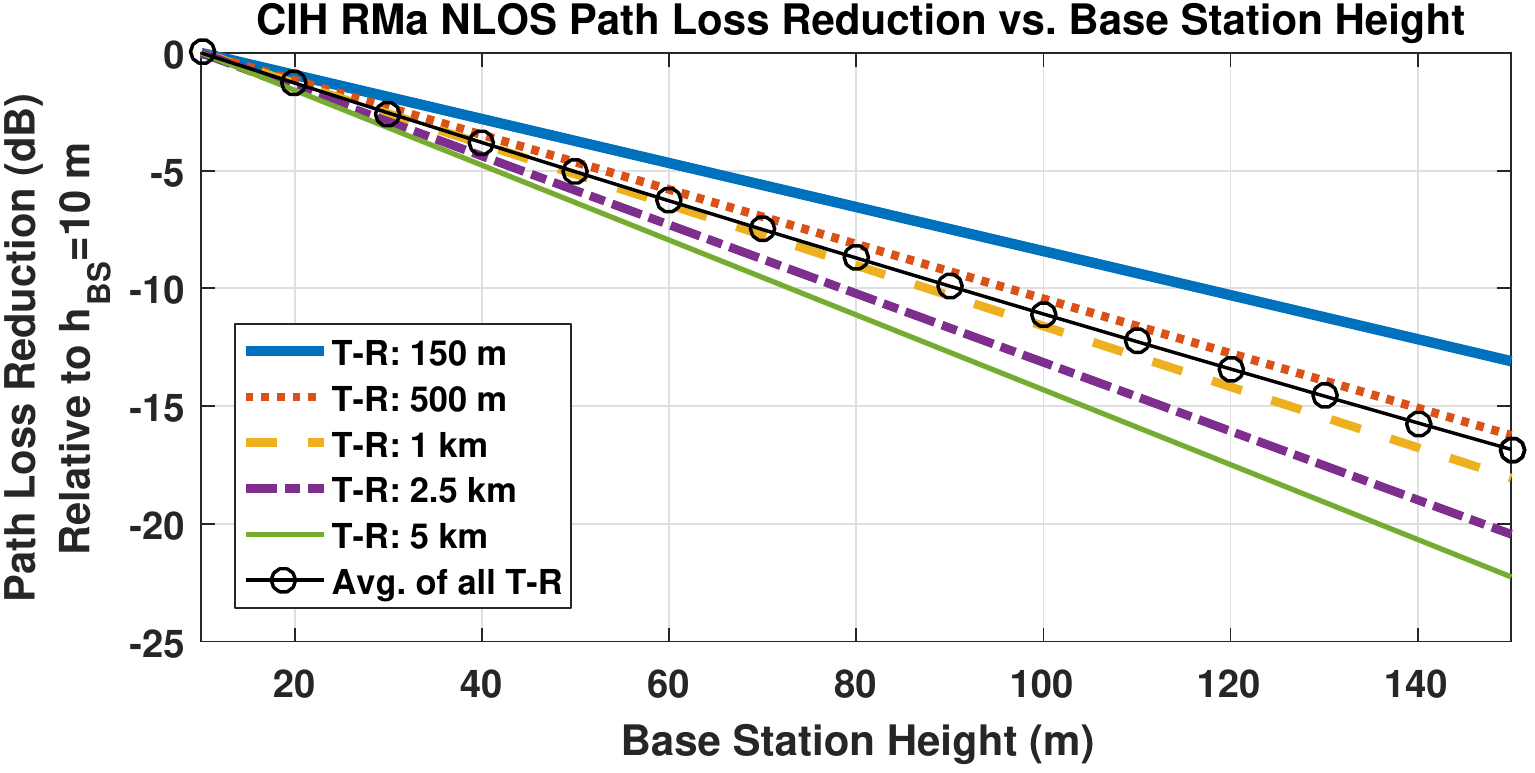}
	\caption{Relationship between TX base station height and decrease in NLOS path loss for T-R separation distances of 150 m, 500 m, 1 km, 2.5 km, and 5 km, and the average over all T-R distances, for the CIH RMa NLOS path loss model in~\eqref{eq:CIHNLOSmeas}. The mobile height ($h_{UT}$) is 1.5 m.}\label{fig:h_BS_CIH_NLOS}	
\end{figure}
\begin{table}
	\centering
	\caption{CI and CIH path loss model parameters from simulations and measurements, and the corresponding equation numbers for each model. The reference base station height for the CIH model is $h_{B0}=35$ m.}\label{tbl:params}
	\scalebox{0.79}{
		\fontsize{9}{9}\selectfont
		\begin{tabular}{|c|c|c|c|c|c|c|}  \specialrule{1.5pt}{0pt}{0pt}
			\multicolumn{7}{|c|}{\textbf{CI and CIH RMa Path Loss Model Parameters}} \\ \specialrule{1.5pt}{0pt}{0pt}
			Model	& Data	& Env.	& Eq.	& \multicolumn{2}{c|}{$\PLE$}	& $\sigma$	\\ \specialrule{1.5pt}{0pt}{0pt}
			$\PL_{\LOS}^{\CI\text{-3GPP}}$		& Sim.	& LOS	& \eqref{eq:CILOS3GPP}	& \multicolumn{2}{c|}{2.31}		& 5.9 dB	\\ \hline
			$\PL_{\LOS}^{\CI\text{-RMa}}$ 		& Meas.	& LOS	& \eqref{eq:CILOSmeas}	& \multicolumn{2}{c|}{2.16}		& 1.7 dB	\\ \hline
			$\PL_{\NLOS}^{\CI\text{-3GPP}}$	& Sim.	& NLOS	& \eqref{eq:CINLOS3GPP}	& \multicolumn{2}{c|}{3.04}		& 8.2 dB	\\ \hline
			$\PL_{\NLOS}^{\CI\text{-RMa}}$ 	& Meas.	& NLOS	& \eqref{eq:CINLOSmeas}	& \multicolumn{2}{c|}{2.75}		& 6.7 dB	\\ \specialrule{1.5pt}{0pt}{0pt}
			Model								& Data	& Env.	& Eq.		& $n$		& $b_{tx}$						& $\sigma$	\\ \specialrule{1.5pt}{0pt}{0pt}
			$\PL_{\LOS}^{\CIH\text{-3GPP}}$	& Sim.	& LOS	& \eqref{eq:CIH3GPP_LOSsim}	& 2.31		& -0.006			& 5.6 dB	\\ \hline
			$\PL_{\LOS}^{\CIH\text{-RMa}}$		& Meas.	& LOS	& \eqref{eq:CIHLOSmeas}	& 2.31		& -0.03				& 1.7 dB	\\ \hline
			$\PL_{\NLOS}^{\CIH\text{-3GPP}}$	& Sim.	& NLOS	& \eqref{eq:CIH3GPP_NLOSsim}	& 3.07		& -0.06				& 8.7 dB	\\ \hline
			$\PL_{\NLOS}^{\CIH\text{-RMa}}$	& Meas.	& NLOS	& \eqref{eq:CIHNLOSmeas}	& 3.07		& -0.049			& 6.7 dB	\\ \specialrule{1.5pt}{0pt}{0pt}
		\end{tabular}}
	\end{table}

The effect of TX height on RMa path loss with the CIH model is evident in~\eqref{eq:CIHLOSmeas}, where the RMa LOS effective PLE reduces from 2.4 to 2.1 when the TX height ranges from 10 m to 150 m. Similarly for NLOS,~\eqref{eq:CIHNLOSmeas} indicates that the RMa effective PLE reduces from 3.2 with a TX height of 10 m, to about 2.6 with a TX height of 150 m. For RMa mmWave propagation, this difference can have an appreciable significance in weather events, where 25 mm/hr rainfall can result in 10 dB loss per km at 73 GHz~\cite{Rap13a}. Fig.~\ref{fig:h_BS_CIH_NLOS} shows the decrease in path loss for various T-R separation distances and corresponding base station heights (similar to Fig.~\ref{fig:3GPPTXheight} for~\eqref{eq:RMaNLOS}) for the CIH NLOS model in~\eqref{eq:CIHNLOSmeas}. This shows that for large T-R separation distances like 5 km, path loss can be reduced by up to 22 dB for a TX height of 150 m compared to 10 m. It is evident that the CIH models in~\eqref{eq:CIHLOSmeas} and~\eqref{eq:CIHNLOSmeas} accurately preserve the effective PLE dependency of TX height in the RMa scenario.

\section{Conclusions}\label{sec:conc}
This paper presented historical details of the 3GPP~\cite{3GPP.38.900} LOS and NLOS RMa path loss models. A key mathematical inconsistency in the 3GPP RMa LOS model~\eqref{eq:RMaLOS} was shown for frequencies above 9.1 GHz when using default model parameters~\cite{3GPP.38.900}. Numerous physical correction factors were described for the 3GPP LOS and NLOS RMa models, such as street width and average building height, which do not make sense for RMa path loss modeling~\cite{Mac16c}. Furthermore, it was identified that the current 3GPP LOS and NLOS RMa path loss models were derived from low-GHz measurements in metropolitan Tokyo, and were not previously verified above 6 GHz or at mmWave bands in a rural setting.

Since little work existed in the literature to verify RMa path loss models at mmWave, we conducted measurements at 73 GHz in rural Virginia~\cite{Mac16c} and derived physically-based CI and CIH path loss models that were verified and shown to be very easy to use as replacements for 3GPP/ITU-R RMa models, without the need for extraneous, nonsensical correction factors. The novel CIH model incorporates path loss dependency on base station height, and best-fit parameters were determined from a combination of simulated and empirical data. The CIH model's use of TX height is a simple, yet powerful option for RMa path loss modeling above 6 GHz.

The CIH RMa LOS and NLOS path loss models derived in~\eqref{eq:CIHLOSmeas} and~\eqref{eq:CIHNLOSmeas}, respectively, indicate that the effective PLE reduces as the TX height increases ($b_{tx}=-0.03$ in LOS; $b_{tx}=-0.049$ in NLOS). When considering a base station height of 150 m compared to 10 m, the average decrease in NLOS path loss across all T-R separation distances derived from the measurements  is 17 dB, whereas the simulated 3GPP model results in a 29 dB decrease in path loss. This could lead to overestimating interference and coverage when using the 3GPP RMa model compared to the CIH model.
 
Finally, this paper suggests that the CI and CIH RMa models proposed in~\eqref{eq:CILOSmeas},~\eqref{eq:CINLOSmeas},~\eqref{eq:CIHLOSmeas}, and~\eqref{eq:CIHNLOSmeas}, should be considered for RMa model adoption in 3GPP and ITU-R for frequencies above 6 GHz~\cite{3GPP.38.900}. While 3GPP specifies the RMa path loss model applicability up to 30 GHz, the goal of TR 38.900 is to provide channel models for frequencies up to 100 GHz. The CI and CIH models herein offer an alternative 3GPP/ITU-R RMa path loss model that can reliably model path loss across the low microwave bands and all the way up through the mmWave bands, beyond 100 GHz, similar to the optional UMa, UMi, and InH path loss models in 3GPP TR 38.900~\cite{3GPP.38.900}. The models herein are validated with real-world mmWave measurements, and have the same mathematical form as the optional path loss model already in 3GPP, and are proven to offer superior prediction accuracy when applied to new frequencies, distances, or use cases~\cite{Sun16b}. Additional measurements across more frequencies and TX heights are needed for further validation. The RMa CI models derived from empirical data are provided in the NYU WIRELESS NYUSIM software that may be used to simulate RMa path loss and channel characteristics for various parameters~\cite{Sun17b}. 

\bibliography{MacCartney_Bibv5}
\bibliographystyle{IEEEtran}

\end{document}